\documentclass[twocolumn,showpacs,preprintnumbers,amsmath,amssymb]{revtex4}
\usepackage{graphicx}
\usepackage{dcolumn}
\usepackage{bm}

\begin{document} 

\title{Second-harmonic generation and linear electro-optical coefficients of SiC polytypes and nanotubes}  
\author{I.\ J.\ Wu and G.\ Y.\ Guo\footnote{Electronic address: 
gyguo@phys.ntu.edu.tw}}
\address{Department of Physics and Center for Theoretical Sciences,
 National Taiwan University, Taipei, Taiwan 106, Republic of China}
\date{\today}

\begin{abstract}
The second-order nonlinear optical susceptibility ($\chi_{abc}^{(2)}$) and linear electro-optical 
coefficient ($r_{abc}$) of a large number of single-walled zigzag, armchair 
and chiral SiC nanotubes (SiC-NTs) as well as bulk SiC polytypes (2H-, 4H-, 6H- and 3C-SiC)
and single graphitic SiC sheet have been calculated from first-principles. 
The calculations are based on density functional
theory in the local density approximation and highly accurate 
full-potential projector augmented-wave method is used.
Both the zigzag and chiral SiC-NTs are found to
exhibit large second-order nonlinear optical behavior with the $\chi_{abc}^{(2)}$
and $r_{abc}$ coefficients being up to ten-times larger than that of 
bulk SiC polytypes, and also being up to thirteen-times larger than the counterparts
of the corresponding BN-NTs, indicating that SiC-NTs are promising materials 
for nonlinear optical and opto-electric applications.
The prominant features in the spectra of $\chi_{abc}^{(2)}(-2\omega,\omega,\omega)$ 
of the SiC-NTs are correlated with the features in the 
linear optical dielectric function $\varepsilon (\omega)$ in terms of single-photon
and two-photon resonances.

\end{abstract}

\pacs{42.65.-k,42.70.Nq,73.22.-f,78.67.Ch}

\maketitle

\section{Introduction}

Carbon nanotubes (CNTs) have attracted considerable interest worldwide 
ever since their discovery in 1991~\cite{iij91},
mainly because of their unusual properties and great 
potentials for technological applications. 
CNTs can be regarded as a layer of graphene sheet rolled up in a tubular
form~\cite{Saito}, and the structure of a CNT is specified by a chiral vector defined
by a pair of integers ($n,m$). CNTs can be chiral or
nonchiral depending on the way they are rolled up.
Their physical properties, in particular, optical dielectric functions, depend sensitively
on their chirality, i.e., the ($n, m$) indices (see, e.g., Refs. \onlinecite{guo04a} 
and \onlinecite{guo04b} and references therein). Apart from CNTs, inorganic tubular materials,
such as BN~\cite{rub94,cho94}, AlN~\cite{Mei}, GaN~\cite{Lee}, have
also been predicted and synthesized. These tubular materials also display some very interesting
properties distinctly different from their bulks.

Bulk silicon carbide (SiC) crystallizes in either a cubic or a hexagonal form, and
exbihits polytypism~\cite{mad82,wyc63}. The polytypes are made of identical hexagonal layers with
different stacking sequences. These polytypes are semiconductors with a range of band gaps,
from 2.39 eV in the zincblende (3C) to 3.33 eV in the wurtzite polytype (2H)~\cite{mad82,iva92}.
Furthermore, 3C- and 6H-SiC are used for high temperature, high-power and high-frequency
devices~\cite{C,Wang,Rurali,Park} due to their unique properties~\cite{SiC}, while 6H-SiC
with a band gap of 2.86 eV is a useful material for blue light-emitting diode applications~\cite{iva92}.
Recently, SiC-NTs were also synthesized by the reaction of CNTs and SiO
at different temperatures~\cite{sun02}.
This has stimulated a number of theoretical and experimental
investigations on the tubular form of the SiC~\cite{pei06,Pei,Menon,Xia,Mavrandonakis}.
Based on density-functional calculations, Miyamoto and Yu~\cite{Miyamoto}
predicted the existence of graphitic and tubular forms of SiC and also proposed
their synthesis using an extreme hole injection technique. They also reported that
the strain energies of SiC-NTs are lower than that of CNTs, and that the band
gaps of SiC-NTs can be direct or indirect, depending on the chirality.
Using both tight-binding molecular dynamics and $\textit{ab initio}$
methods, Menon and coworkers~\cite{Menon} showed that single-walled SiC-NTs are highly stable
with a large band gap.  Zhao $\emph{et al.}$~\cite{Zhao}
also investigated theoretically the strain energy, atomic and electronic structure of SiC-NTs with
or without hydrogenation. Gali performed an {\it ab initio} study of the effect of
nitrogen and boron impurities on the band structure of the SiC-NTs~\cite{Gali}.

Unlike CNTs, SiC-NTs are polar materials and therefore, may exhibit
some unusual physical properties that CNTs may not have. For example, 
like BN-NTs,\cite{nak03,Lin}
zigzag SiC-NTs may become piezoelectric, and also show second-order
non-linear optical response.
Despite the intensive theoretical studies mentioned above,
no $\textit{ab initio}$ calculation of the dielectric response and optical
properties of SiC-NTs has been reported, perhaps because of the heavy
demand on computer resources. 
A knowledge of the optical properties of SiC-NTs
is important for their optical and electrooptical applications.
Therefore, we have recently carried out a series of {\it ab initio} calculations 
in order to analyze the linear optical features and underlying band structure
of all three types of the SiC-NTs as well as their possible dependence on diameter 
and chirality.\cite{ijwu}
In this work, we investigate the second-order optical susceptibility and also linear
electro-optical coefficient of the SiC-NTs as well as bulk SiC polytypes.
The primary objective of this work is to
find out the features and magnitude of the second-harmonic generation and linear 
electro-optical coefficients of the SiC-NTs in order to see whether they have
any potential applications in nonlinear optical and electro-optical devices such as
second-harmonic generation, sum-frequency generation, and electrical optical switch.
The second objective is to identify distinguished differences in nonlinear optical
properties between SiC-NTs and CNTs~\cite{guo04a}, and also between SiC-NTs and BN-NTs~\cite{Lin}.

The rest of this paper is organized as follows. In Sec. II, the theoretical approach and
computational method are described. In Sec. III, the calculated second-order nonlinear
optical susceptibility and linear electro-optical coefficients of SiC nanotubes are
presented. For comparison, the second-order nonlinear optical susceptibility 
and linear electro-optical coefficients of bulk SiC polytypes and single graphitic SiC sheet
are also calculated and presented.
Finally, in Sec. IV, a summary is given.

\section{Theory and computational method}

Our {\it ab initio} calculations for the SiC-NTs were performed using
highly accurate full-potential projector augmented-wave (PAW)
method~\cite{blo94}, as implemented in the vasp package~\cite{kre93}.
They are based on density functional theory (DFT) with the local density 
approximation (LDA).
A supercell geometry was adopted so that the nanotubes are aligned in a
square array with the closest distance between adjacent nanotubes being
at least 10 \AA. A large plane-wave energy cutoff of 450 eV was used throughout.
We consider a large number of representative SiC-NTs with a range of
diameters from all three types, as listed in Ref. \onlinecite{ijwu}. 

First, the ideal nanotubes were constructed by rolling-up a 
hexagonal SiC sheet. 
The atomic positions and lattice constants were then fully relaxed by the conjugate 
gradient technique. The theoretical equilibrium atomic positions and lattice
constant were obtained when the forces acting on all the atoms and the uniaxial 
stress were less than 0.04 eV/\AA$ $ and 1.0 kBar, respectively.
The calculated equilibrium lattice constants $T$
and curvature energies $E_c$ (total energy relative to that of
single SiC sheet) as well as the other computational details have been 
reported before~\cite{ijwu}.

The final self-consistent electronic band structure calculations were then
carried out for the theoretically determined SiC-NT structures. 
In this work, the non-linear optical properties were calculated based on the
independent-particle approximation, i.e., the excitonic effects and the 
local-field corrections were neglected. Note that the band gap of SiC 
in the wurtzite (2H) structure
from our LDA calculations is 2.12 eV (see Sec. IIIA below), being 1.18 eV smaller than
the experimental value~\cite{mad82,iva92}. This suggests that the quasi-particle
self-energy corrections to the optical peak positions may amount to $\sim$1 eV
in the SiC systems. Furthermore, because the LDA underestimates the energy gaps, the
calculated dielectric function and second-order nonlinear optical susceptibility
in the static limit might be slightly too large in general. For example,
our calculated dielectric function at 0 eV for the electric field parallel 
and perpendicular
to the $c$-axis of 2H-SiC is 8.33 and 7.90, respectively, being slightly larger
than the corresponding measured $\epsilon_{\infty}$, 6.84 and 6.51.~\cite{mad82}
Therefore, the results from the present LDA-independent-particle calculations
might not be able to be compared quantitatively with experiments, though they
would certainly be useful to study the trends and characteristics of the
optical properties of the SiC-NTs.
Nonetheless, our previous calculations show that the dielectric functions of
graphite~\cite{guo04a} and also of h-BN~\cite{Lin} calculated within 
the independent-particle picture are in reasonable agreement with experiments, 
perhaps because of the accidental
cancellation of the self-energy corrections by the excitonic effects.

Following previous nonlinear optical calculations~\cite{dua99,guo04a,Lin}, the imaginary
part of the second-order
optical susceptibility due to direction interband transitons is given by~\cite{gha90}
\begin{equation}
 \chi^{''(2)}_{abc}(-2\omega,\omega,\omega) = \chi^{''(2)}_{abc,VE}(-2\omega,\omega,\omega)
 + \chi^{''(2)}_{abc,VH}(-2\omega,\omega,\omega) 
\end{equation}
where the contribution due to the so-called virtual-electron (VE) process is
\begin{eqnarray} \label {eq:VE}
\chi^{''(2)}_{abc,VE} = -\frac{\pi}{2\Omega}\sum_{i\in VB} \sum_{j,l\in CB}
 \sum_{\bf k} w_{\bf k}
 \{ \frac{Im[p_{jl}^a\langle p_{li}^bp_{ij}^c\rangle]}{\epsilon^3_{li}(\epsilon_{li}+
 \epsilon_{ji})}\delta(\epsilon_{li}-\omega) \nonumber \\
 -\frac{Im[p_{ij}^a\langle p_{jl}^bp_{li}^c\rangle]}{\epsilon^3_{li}(2\epsilon_{li}-
  \epsilon_{ji})}\delta(\epsilon_{li}-\omega) 
  +\frac{16Im[p_{ij}^a\langle p_{jl}^bp_{li}^c\rangle]}{\epsilon^3_{ji}(2\epsilon_{li}-
  \epsilon_{ji})}\delta(\epsilon_{ji}-2\omega) \}
\end{eqnarray}
and that due to the virtual-hole (VH) process
\begin{eqnarray} \label {eq:VH}
\chi^{''(2)}_{abc,VH} = \frac{\pi}{2\Omega}\sum_{i,l\in VB} \sum_{j\in CB}
 \sum_{\bf k} w_{\bf k}
 \{ \frac{Im[p_{li}^a\langle p_{ij}^bp_{jl}^c\rangle]}{\epsilon^3_{jl}(\epsilon_{jl}+
 \epsilon_{ji})}\delta(\epsilon_{jl}-\omega) \nonumber \\
 -\frac{Im[p_{ij}^a\langle p_{jl}^bp_{li}^c\rangle]}{\epsilon^3_{jl}(2\epsilon_{jl}-
 \epsilon_{ji})}\delta(\epsilon_{jl}-\omega) 
  +\frac{16Im[p_{ij}^a \langle p_{jl}^bp_{li}^c\rangle]}{\epsilon^3_{ji}(2\epsilon_{jl}-
 \epsilon_{ji})}\delta(\epsilon_{ji}-2\omega) \}.
\end{eqnarray}
Here $\epsilon_{ji} = \epsilon_{{\bf k}j}-\epsilon_{{\bf k}i}$ and
$\langle p_{jl}^bp_{li}^c\rangle = \frac{1}{2}(p_{jl}^bp_{li}^c+p_{li}^bp_{jl}^c)$.
The dipole transition matrix elements $p_{ij}^a = <{\bf k}j|\hat{p}_a|{\bf k}i>$ were obtained from
the self-consistent band structures within the PAW formalism ~\cite{ado01}.
The real part of the second-order optical susceptibility is then obtained from
$\chi''^{(2)}_{abc}$ by
a Kramer-Kronig transformation
\begin{equation}
\chi'^{(2)}(-2\omega,\omega,\omega) = \frac{2}{\pi}{\bf P} \int_0^{\infty}d\omega'
 \frac{\omega'\chi''^{(2)}(2\omega',\omega',\omega')}{\omega'^2-\omega^2}.
\end{equation}

The linear electro-optic coefficient $r_{abc}(\omega)$ is connected to the 
second-order optical susceptibility $\chi^{(2)}_{abc}(-\omega,\omega,0)$ through 
the relation~\cite{hug96}
\begin{equation}
\chi^{(2)}_{abc}(-\omega,\omega,0) = -\frac{1}{2} n^2_{a}(\omega)n^2_{b}(\omega)r_{abc}(\omega)
\end{equation}
where $n_a(\omega)$ is the refraction index in the $a$-direction. In the zero frequency limit,
\begin{equation}
\lim_{\omega\rightarrow 0}\chi^{(2)}_{abc}(-2\omega,\omega,\omega) 
=\lim_{\omega\rightarrow 0}\chi^{(2)}_{abc}(-\omega,\omega,0).
\end{equation}
Therefore, 
\begin{equation}
r_{abc}(0) = -\frac{2}{n^2_{a}(0)n^2_{b}(0)}\lim_{\omega\rightarrow 0}\chi^{(2)}_{abc}(-2\omega,\omega,\omega).
\end{equation}
Furthermore, for the photon energy $\hbar\omega$ well below the band gap,
the linear electro-optic coefficient $r_{abc}(\omega) \approx r_{abc}(0)$
because $\chi^{(2)}_{abc}(-2\omega,\omega,\omega)$ and $n(\omega)$ 
are nearly constant in this low frequency region, as shown in the next section and 
in Ref. ~\cite{ijwu}.
 
In the present calculations, the $\delta$-function in Eqs. (2)-(3)
is approximated by a Gaussian function with $\Gamma = 0.2$ eV.
A uniform $k$-point grid (1$\times$1$\times$n) along the nanotube 
axis ($z$-axis) with the number $n$ being from 40 to 100, is used.
Furthermore, to ensure that $\chi'^{(2)}_{abc}$ calculated via Kramer-Kronig
transformation (Eq. (4)) is reliable, at least ten energy bands 
per atom are included in the present optical calculations. The unit cell
volume $\Omega$ in Eqs. (2)-(3) is not well defined for nanotubes.
Therefore, like the previous calculations~\cite{guo04a,guo04b,Lin,ijwu}, we used the
effective unit cell volume of the nanotubes rather than the volume of the supercells
which is arbitrary. The effective unit cell of a nanotube is given by
$\Omega = \pi[(D/2+d/2)^2-(D/2-d/2)^2]T = \pi DdT$ where $d$ is the thickness of the
nanotube cylinder which is set to the interlayer distance of $h$-SiC 
(3.51 \AA~\cite{ijwu}). $D$ and $T$ are the diameter and length of 
translational vector of the nanotube~\cite{ijwu}, respectively.  

\section{Results and discussion}

\begin{table*}
\caption{Calculated lattice constants of SiC polytypes, together with other
theoretical (Cal.) (Bechstedt {\it et al.}~\cite{bechstedt}) and experimental (Exp.) 
(Ref. \onlinecite{mad82}) lattice constants.}
\begin{ruledtabular}
\begin{tabular}{ c c c c c }
              & 2H    & 4H     &  6H    & 3C   \\
              & a (\AA), c (\AA) & a (\AA), c (\AA) & a (\AA), c (\AA)  & a (\AA) \\  \hline
This work &3.057, 5.018 &3.060, 10.018 & 3.061, 15.018 &4.331 \\
Cal. &3.057, 5.016 &3.061, 10.012 &3.062, 15.012 &4.332 \\
Exp. &3.076, 5.048 &3.073, 10.053 &3.081, 15.117 &4.360 \\
\end{tabular}
\end{ruledtabular}
\end{table*}

\subsection{Bulk SiC polytypes}

For comparison with the SiC-NTs, we first investigated the second-order 
nonlinear optical susceptibility of bulk SiC polytypes.
SiC exists in either a cubic or a hexagonal polymorphic structure. 
The polytypes are made of identical hexagonal layers with different stacking 
sequences. SiC occurs in nearly 200 polytypes.\cite{jep83} 
Here we consider only the four common 2H, 4H, 6H and 3C types of SiC.
The zincblende (3C) structure, with pure cubic stacking of the Si-C double layers in
the [111] direction, is one of the two most extreme polytypes. The other is the
wurtzite (2H) structure with pure hexagonal stacking in the [0001] direction. 
The atomic positions and lattice constants of the four 
polytypes were theoretically optimized.
In the atomic structure optimizations, a uniform $k$-point grid of $30\times30\times n$ 
with $n$ being from 15 to 30 was used. 
The calculated equilibrium lattice constants of the SiC polytypes are 
listed in Table I, together with previous theoretical and experimental
values. Table I shows that our theoretical lattice constants are in very good
agreement with the previous calculations~\cite{bechstedt}, though they are
slightly smaller (within 0.7 \%) than the corresponding experimental 
values~\cite{mad82}.
Our calculated indirect band gaps for the 3C, 2H, 4H and 6H types of SiC
are, respectively, 1.29 eV, 2.12 eV, 2.17 eV and 1.96 eV. These band gap values
are smaller by about 1.1 eV than the corresponding experimental values~\cite{mad82}
of 2.39 (3C), 3.33 eV (2H), 3.26 eV (4H), and 2.86 eV (6H).  
Nonetheless, our theoretical band gap values are in very good agreement with
previous LDA calculations (see, e.g., Ref. \onlinecite{bernd} and references therein).
The difference in the theoretical DFT-LDA and experimental band gap values 
can be attributed to the quasi-particle self-energy corrections, as has been
shown explicitly in previous quasiparticle band structure calculations for 
the SiC polytypes~\cite{bernd}.

\begin{figure*}
\includegraphics[width=14cm]{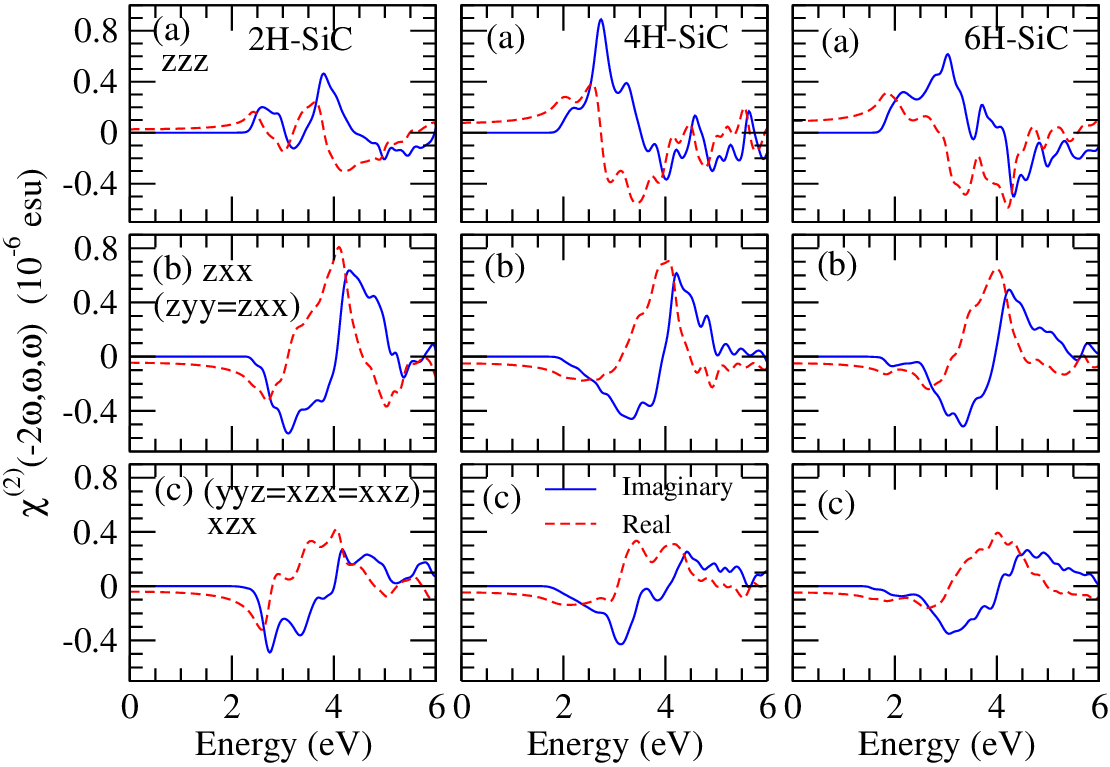}
\caption{\label{fig1}(Color online) Real and imaginary parts of
$\chi^{(2)}(-2\omega,\omega,\omega)$ of the hexagonal SiC polytypes.}
\end{figure*}

In Fig. 1, we display the imaginary and real parts of three nonvanishing components of the
second-order nonlinear optical susceptibility (tensor) $\chi^{(2)}(-2\omega,\omega,\omega)$ 
of the three hexagonal SiC polytypes. The imaginary and real parts of the nonvanishing 
component $\chi_{xyz}^{(2)}(-2\omega,\omega,\omega)$ of the cubic SiC are plotted in Fig. 2. 
In the nonlinear optical calculations,
we used a denser $k$-point grid of $50\times50\times\ n$ with $n$
being from 10 to 30 for the three hexagonal SiC polytypes, and $n$ = 50 for the cubic SiC.
Importantly, we find numerically that only the $\chi^{(2)}_{zzz}$,
$\chi^{(2)}_{zxx}$, $\chi^{(2)}_{zyy}$, $\chi^{(2)}_{xzx}$, $\chi^{(2)}_{yzy}$, $\chi^{(2)}_{xxz}$,
 and $\chi^{(2)}_{yyz}$ components are nonzero for the three hexagonal polytypes.
Moreoever, $\chi^{(2)}_{zyy} = \chi^{(2)}_{zxx} $, 
and $\chi^{(2)}_{yyz} = \chi^{(2)}_{xxz} = \chi^{(2)}_{xzx} = \chi^{(2)}_{yzy}$. 
This is consistent
with the symmetry consideration,~\cite{boy} demonstrating that our
numerical method and calculations are qualitatively correct.
For the 3C-SiC, only $\chi^{(2)}_{xyz}$, $\chi^{(2)}_{zxy}$, $\chi^{(2)}_{yzx}$ components
are nonzero, and all the three components are numerically found to be identical,
as they should. Fig. 1 shows that the line shape, amplitude and feature positions of
the $\chi^{(2)}_{zxx}$ and $\chi^{(2)}_{xzx}$ components for all the hexagonal SiC polytypes 
are very similar. 
On the other hand, the $\chi^{(2)}_{zzz}$ component varies significantly as one moves from
2H- to 6H-SiC. This is mainly because in this case all the electric fields are polarized parallel to the
$z$-axis and hence the $\chi^{(2)}_{zzz}$ component should be rather sensitive to the layer stacking
sequence along the $z$-axis. 
We also note that the
shape of the $\chi^{(2)}_{xyz}$ of 3C SiC is similar to that of the $\chi^{(2)}_{zzz}$
of 6H-SiC (Fig. 1 and Fig. 2). Nonetheless, the features in 
the $\chi^{(2)}_{xyz}$ of the cubic structure appears to have the largest magnitude
among the four SiC polytypes studied here.

\begin{figure}
\includegraphics[width=8cm]{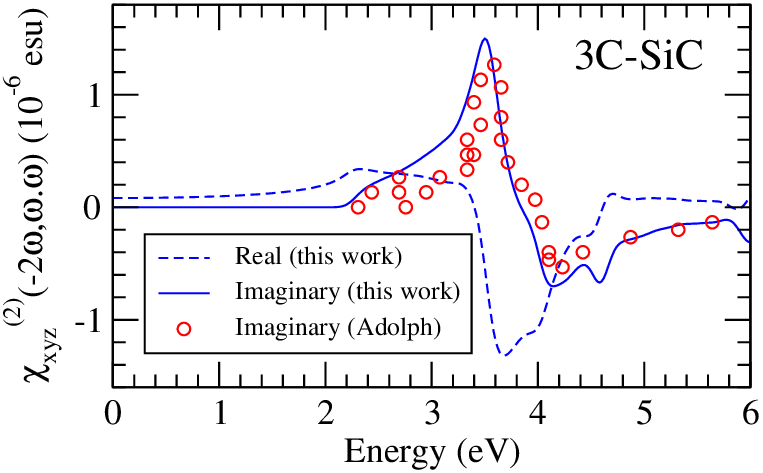}
\caption{\label{fig2} (Color online) Real and imaginary parts
of $\chi_{xyz}^{(2)}(-2\omega,\omega,\omega)$ of SiC in the zinc-blende (3C) structure.
Previous calculations by Adolph {\it et al.}~\cite{Adolph} 
are also displayed for comparison.}
\end{figure}

We find that our calculated nonzero components of $\chi^{(2)}(-2\omega,\omega,\omega)$
of all the four polytypes agree quite well with the recent DFT-LDA calculations by
Adolph and Bechstedt~\cite{Adolph}. For detailed comparison, we plot the imaginary 
part of the nonzero components of SiC in the cubic and wurtzite structures from 
the previous calculations by
Adolph and Bechstedt~\cite{Adolph} 
in Fig. 2 (cubic) and Fig. 3 (wurtzite), together with our results.
One can see from Figs. 2 and 3(c) that $\chi_{xyz}^{(2)}(-2\omega,\omega,\omega)$
of 3C-SiC and $\chi_{xzx}^{(2)}(-2\omega,\omega,\omega)$ of 2H-SiC from the present and 
previous calculations are almost identical. Nonetheless, there are some discernable
discrepancies in the $\chi_{zxx}^{(2)}$ and $\chi_{zzz}^{(2)}$ components of 2H-SiC between
the two calculations [Fig. 3(a-b)]. 

\begin{table}
\caption{Calculated static refraction index $n$, second-order optical
susceptibility $\chi^{(2)}(0)$ (pm/V)  and linear electro-optical
coefficient $r_{abc}$ (pm/V) of the hexagonal SiC polytypes (a), zinc-blende SiC (b)
and also isolated SiC sheet (c). 
}
\begin{ruledtabular}
\begin{tabular}{c c c c }
 (a)    &$n_x$ ($n_z$) & $\chi^{(2)}_{zzz}$, $\chi^{(2)}_{zxx}$
                       & $r_{zzz}$, $r_{zxx}$  \\ \hline
2H-SiC  & 2.81 (2.88)  & 11.5, -17.9 & -0.33, 0.55  \\
4H-SiC  & 2.83 (2.87)  & 32.8, -20.4 & -0.97, 0.62  \\
6H-SiC  & 2.82 (2.89)  & 38.6, -20.4 & -1.11, 0.61  \\
        &              &               &              \\
 (b)    &$n$   & $\chi^{(2)}_{xyz}$ &  $r_{xyz} $  \\ \hline
3C-SiC  & 2.95   &  34.2   &  -0.90  \\
        &              &               &              \\
 (c)    &$n_a$ ($n_c$) & $\chi^{(2)}_{bbb}$, $\chi^{(2)}_{aab}$ & $r_{bbb}$, $r_{aab}$ \\ \hline
SiC sheet  & 3.19 (2.01) & 227.8, -227.8 & -11.08, 11.08    \\
\end{tabular}
\end{ruledtabular}
\end{table}

In Table II, the calculated zero frequency linear electro-optic coefficient
$r(0)$ as well as the corresponding second-order nonlinear optical susceptibility
$\chi^{(2)}(0,0,0)$ are listed. The $r(0)$ is calculated from the
corresponding $\chi^{(2)}(0,0,0)$ by using Eq. (7).
We find that the calculated $\chi^{(2)}_{xzx}(0)$ and $\chi^{(2)}_{zxx}(0)$ for the
three hexagonal SiC polytypes differs slightly from each other, i.e.,
only approximately satisfying the requirement of the
Kleiman symmetry~\cite{kle62}. In Table II, therefore, the $\chi^{(2)}_{zxx}(0)$ 
is listed as the averaged value of $\chi^{(2)}_{xzx}(0)$ and $\chi^{(2)}_{zxx}(0)$,
and hence $r_{zxx}(0)$ as the averaged value of $r_{zxx}(0)$
and $r_{xzx}(0)$, for all the hexagonal SiC polytypes.
Clearly, Table II shows that $\chi^{(2)}_{zxx}(0)$ and hence $r_{zxx}(0)$
are nearly the same for all the hexagonal SiC polytypes.
In contrast, $\chi^{(2)}_{zzz}(0)$ and hence $r_{zzz}(0)$ increases substantially
as one moves from 2H-SiC to 6H-SiC, suggesting that $\chi^{(2)}_{zzz}(0)$ 
and $r_{zzz}(0)$ are more sensitive to the layer stacking sequence along the
$z$-axis. This may be expected since all the electric fields are polarized 
parallel to the $z$-axis in the latter case. 

Our calculated nonvanishing components of $\chi^{(2)}(0)$ for the four polytypes
are compared with previous {\it ab initio} calculations\cite{Adolph,Sergey,chen}
and available experiments\cite{nie,lund} in Table III.
Clearly, our results agree rather well with the previous {\it ab initio} 
calculations by Chen {\it et al.}\cite{chen}. Nonetheless, both the results 
of ours and of chen {\it et al.}\cite{chen} 
appear to be nearly two-times larger than the previous {\it ab initio}
calculations by Adolph {\it et al.}\cite{Adolph} and Rashkeev {\it et al.}\cite{Sergey}.
This is surprising since the imaginary parts of $\chi^{(2)}$ from both the present
and previous \cite{Adolph} calculations are very similar (Figs. 2 and 3).
The precise origin of these discrepancies is not known at moment.
Finally, we would consider that the agreement between our results and the 
experimental values is rather good (Table III), given the large uncertainties 
in the available experimental data.\cite{nie,lund}
 
\begin{figure}
\includegraphics[width=8cm]{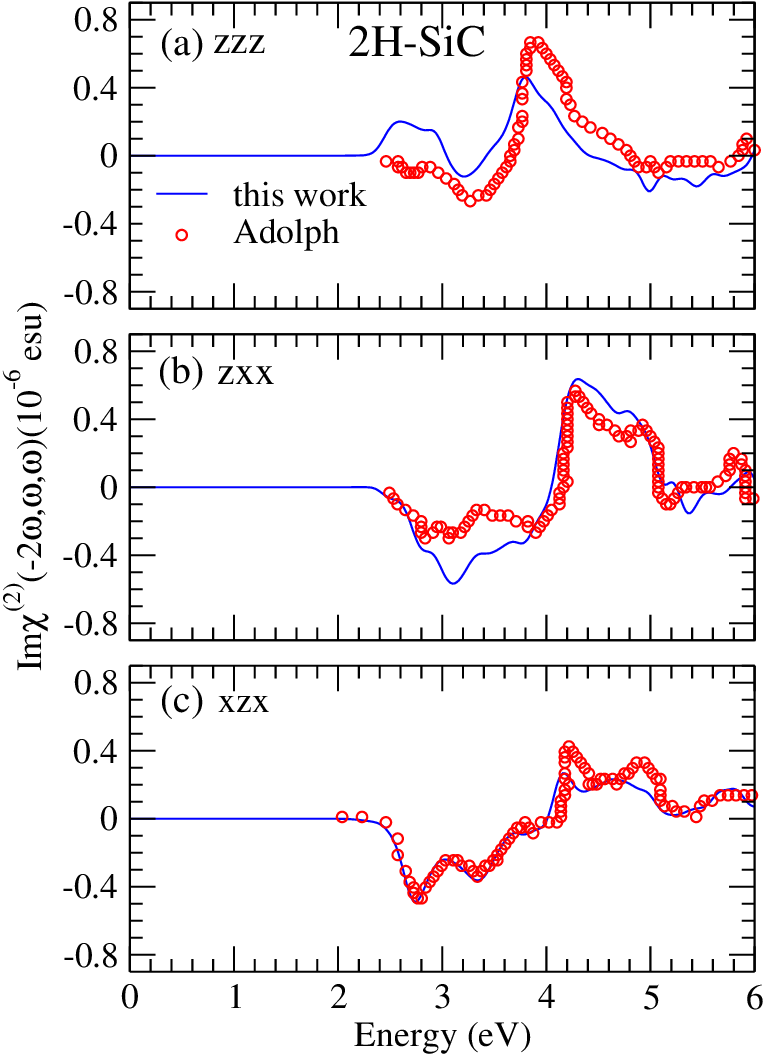}
\caption{\label{fig3} (Color online) Imaginary parts of $\chi^{(2)}(-2\omega,\omega,\omega)$ of 
SiC in the wurtzite (2H) structure.
Previous calculations by Adolph {\it et al.}~\cite{Adolph} 
are also displayed for comparison.}
\end{figure}

\subsection{Single graphitic SiC sheet}

In order to understand the nonlinear optical properties of the SiC-NTs,
we have also calculated the second-order nonlinear optical
susceptibilities for an isolated honeycomb SiC sheet. The isolated SiC
sheet is simulated by a slab-supercell approach with an intersheet distance
of about 10 \AA. The underlying structure was determined theoretically by
using $k$-mesh of $100\times 100\times 1$, and the theoretical lattice constant
is $a$ = 3.069 \AA. Interestingly, we find numerically that the $\chi^{(2)}_{aab}$,
$\chi^{(2)}_{baa}$ and $\chi^{(2)}_{bbb}$ for the isolated SiC sheet
are nonzero. Here $a$ and $b$ denote the two Cartesian coordinates
within the SiC layer. Moreoever, $\chi^{(2)}_{baa} = \chi^{(2)}_{aab} $
and $\chi^{(2)}_{bbb} = -\chi^{(2)}_{aab} $. This is consistent
with the symmetry consideration, demonstrating again that our
numerical method and calculations are qualitatively correct.
It is because that the isolated SiC sheet does not have the spatial
inversion symmetry ($D_{6h}^4$) and its symmetry class
is $D_{3h}$ ($P\bar{6}m2$). Therefore, the isolated
SiC sheet has nonzero $\chi^{(2)}_{aab}$, $\chi^{(2)}_{baa}$ and
$\chi^{(2)}_{bbb}$ from the symmetry consideration.~\cite{boy}

\begin{figure}
\includegraphics[width=8cm]{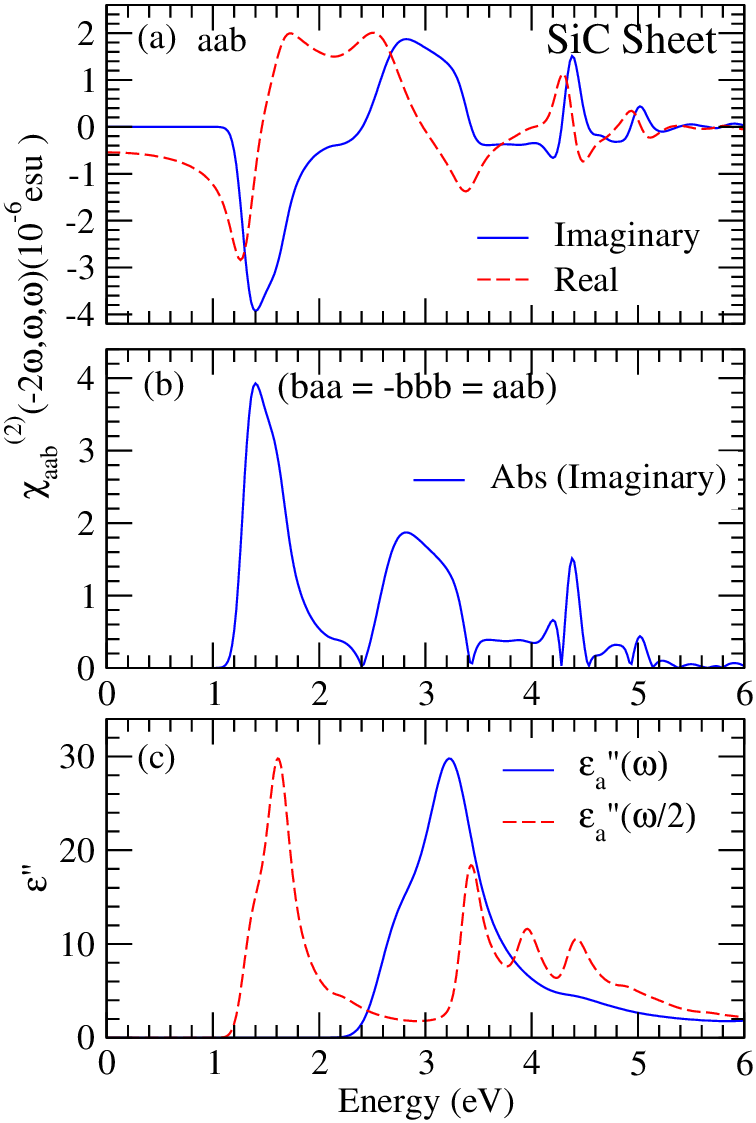}
\caption{\label{fig4} (Color online) (a) Real and imaginary parts as well as (b)
the absolute value of the imaginary part of $\chi^{(2)}_{aab}(-2\omega,\omega,\omega)$
of the isolated SiC sheet. In (c), $\varepsilon_a''(\omega)$
and $\varepsilon_a''(\omega/2)$ (imaginary part of the
dielectric function) from Ref. ~\cite{ijwu} are plotted.}
\end{figure}

We show in Fig. 4 the calculated real and imaginary parts as well as
the absolute value of the imaginary part of $\chi^{(2)}_{aab}(-2\omega,\omega,\omega)$
of the single SiC sheet. 
It is clear from Fig. 4 that the second harmonic generation (SHG) coefficient 
$\chi^{(2)}_{aab}(-2\omega,\omega,\omega)$ is significant in the entire
range of the optical photon energy ($\hbar\omega$).
Furthermore, for the photon energy smaller than about 1.2 eV, the $\chi^{(2)}_{aab}$
is purely dispersive (i.e., real and lossless) [Fig. 4(a)], suggesting that
the SiC sheet has potential application in nonlinear optical devices. 
Note that since the SiC sheet has a theoretical band gap ($E_g$) 
of $\sim$2.6 eV, the absorptive (imaginary) part of the $\chi^{(2)}_{aab}$ 
becomes zero below the half of the band gap (i.e., $\sim$1.2 eV). 
The real part of the $\chi^{(2)}_{aab}$ remains nearly constant at 
low photon energies up to 0.5 eV, then increases steadily 
in magnitude as the photon energy increases, and finally peaks at the
absorption edge of $\sim$1.2 eV [Fig. 4(a)]. In the energy range from 1.4 to 
3.0 eV, the real part of the $\chi^{(2)}_{aab}$ becomes positive and
forms a broad double peak structure. Beyond 3.0 eV, it becomes negative
again and its magnitude gradually deminishes as the photon energy
further increases [Fig. 4(a)].
 
The absorptive part of the $\chi^{(2)}_{aab}$ 
is nonzero only above $\sim$1.2 eV, and looks like a Lorentzian 
oscillation between 1.2 and 3.5 eV with one sharp negative peak at $\sim$ 
1.4 eV and one broad positive peak around 2.8 eV [Fig. 4(a)]. 
It is clear from Eqs. (2) and (3) that the
calculated $\chi^{(2)}_{aab}$ spectra can have pronounced features due
to both single- and double-frequency resonant terms. To
analyze the features in the calculated $\chi^{(2)}$ spectra, it is helpful
to compare the absolute value of $\chi''^{(2)}$ [Fig 4(b)] with the 
absorptive part of the corresponding dielectric function
$\varepsilon''$. Therefore, the calculated $\varepsilon''$
from our previous publication~\cite{ijwu} are shown in Fig. 4(c) as a function
of both $\omega /2$ and $\omega$. Clearly, the first sharp peak at $\sim$1.4 eV
is due to two-photon resonances [cf. $\varepsilon_a''(\omega/2)$] while in contrast,
the second broad peak around 2.8 eV comes from the single-photon resonances
[cf. $\varepsilon_a''(\omega)$]. Nevertheless, both single- and double-photon resonances
involve only interband $\pi \rightarrow \pi^{\ast}$
and $\sigma \rightarrow \sigma^{\ast}$ optical transitions for the
electric field vector ${\bf E}$ polarized parallel to the SiC layer
($E \parallel \hat{a}$)~\cite{ijwu}.

\begin{table*}
\caption{Calculated  second-order optical susceptibility $\chi^{(2)}(0)$
(pm/V) of the four SiC polytypes, together with other theoretical
[Adolph {\it et al.}~\cite{Adolph}, Rashkeev {\it et al.}~\cite{Sergey},
Chen {\it et al.}~\cite{chen}] and experimental [Lundquist {\it et al.}~\cite{lund},
Niedermeier {\it et al.}~\cite{nie}] results.}
\begin{ruledtabular}
\begin{tabular}{ c c c c c }
              & 2H    & 4H     &  6H    & 3C   \\ 
              & $\chi^{(2)}_{zzz}$, $\chi^{(2)}_{zxx}$ 
              & $\chi^{(2)}_{zzz}$, $\chi^{(2)}_{zxx}$ 
              & $\chi^{(2)}_{zzz}$, $\chi^{(2)}_{zxx}$ 
              & $\chi^{(2)}_{xyz}$ \\  \hline            
This work      &11.5,-17.9 &32.8,-20.4 & 38.6,-20.4  &34.2       \\ \hline
Adolph {\it et al.}~\cite{Adolph}   &4.5, -7.1    &15.5, -9.7  &18.1, -9.8  &17.7  \\
Rashkeev {\it et al.}~\cite{Sergey} &3.6, -6.1    &14.5, -8.9  &17.8, -9.7  &17.5  \\
Chen {\it et al.}~\cite{chen}       &8.6, -13.2   &23.3, -14.8 &27.6, -15.0 &24.4  \\ \hline
Lundquist {\it et al.}~\cite{lund}  &   &            &86, -8.6    &17.3  \\
Niedermeier {\it et al.}~\cite{nie} &   &18$\pm$8, -4$\pm$2  &24$\pm$10, -4$\pm$2 &     \\
\end{tabular}
\end{ruledtabular}
\end{table*}

In Table II, the calculated zero frequency linear electro-optic coefficient
$r(0)$ as well as the corresponding second-order nonlinear optical 
susceptibility $\chi^{(2)}(0,0,0)$ and the static refraction
index $n(0)$ are listed. The refraction index $n(0) (=\sqrt{\varepsilon(0)})$
is derived from the calculated static dielectric constant
$\varepsilon(0)$ which has been reported in our recent publication~\cite{ijwu}.
Significantly, the linear electro-optical coefficients of the isolated SiC sheet are 
about ten times larger than that of the bulk SiC polytypes (Table II). 
The static second-order optical susceptibility for the 
SiC sheet is nearly six times larger than the largest component of $\chi^{(2)}(0,0,0)$
of the SiC polytypes.  
Furthermore, the zero frequency linear electro-optic coefficient
$r(0)$ and the corresponding second-order nonlinear optical
susceptibility $\chi^{(2)}(0,0,0)$ of the SiC sheet are more than five times
larger than the corresponding values of the BN sheet~\cite{Lin}. 

\subsection{Second-order optical susceptibility}

We have explicitly calculated the second-order optical susceptibility for the
zigzag (5,0), (6,0), (8,0), (9,0), (12,0), (16,0), (20,0),
(24,0), armchair (3,3), (4,4), (5,5), (8,8), (12,12), (15,15),
and chiral (4,2), (6,2), (8,4), (10,4) SiC-NTs.
In the case of CNTs, only the chiral nanotubes would
exhibit second-order nonlinear optical behavior with
two nonvanishing components of $xyz$ and $xzy$ of $\chi^{(2)}$.~\cite{guo04a}
Here $z$ refers to the coordinate along the tube axis while
$x$ and $y$ denote the two coordinates that are perpendicular to the
tube axis. As for CNTs, the armchair SiC-NTs are found not to have any nonzero components
of $\chi^{(2)}$. 
In contrast, as in the case of the BN-NTs~\cite{Lin}, both the zigzag and chiral 
SiC-NTs are found to show second-order nonlinear optical behavior.
Specifically, all the zigzag SiC-NTs except (5,0),and (9,0), have six nonvanishing
components of the second-order optical susceptibility, namely,
$xzx$, $xxz$, $yyz$, $zxx$, $zyy$, $zzz$. Nevertheless, these components are not completely 
independent of each other. In particular,
$\chi^{(2)}_{xxz}= \chi^{(2)}_{yyz} = \chi^{(2)}_{xzx}$, and 
$\chi^{(2)}_{zyy} = \chi^{(2)}_{zxx}$.
The chiral SiC-NTs have eight nonvanishing
components of the second-order optical susceptibility with two additional nonzero
components being $xyz$ and $yzx$. Note that $\chi^{(2)}_{yzx} = -\chi^{(2)}_{xyz}$.
These numerical findings are consistent with the consideration of the symmetry of
the SiC nanotubes. The point symmetry groups of the SiC-NTs~\cite{boy} are 
$C_{2nv}$ for zigzag ($n$,0) nanotubes, $C_{2nh}$ for armchair ($n,n$) nanotubes, and
$C_N$ for chiral ($n,m$) nanotubes where $N$ = $2(n^2+m^2+nm)/d_R$ with $d_R$ being
the greatest common divisor of $2n+m$ and $2m+n$. Therefore, these symmetries would
dictate~\cite{boy} that all the components vanish for armchair ($n,n$) nanotubes, and that
nonvanishing components for zigzag ($n$,0) nanotubes are $xzx$ = $yzy$, $xxz$ = $yyz$,
$zxx = zyy$, $zzz$, as well as that nonvanishing components of chiral ($n,m$) nanotubes 
include all that for zigzag ($n$,0) nanotubes plus $xyz$ = -$yzx$.  

\begin{figure*}[tb]
\includegraphics[width=16cm]{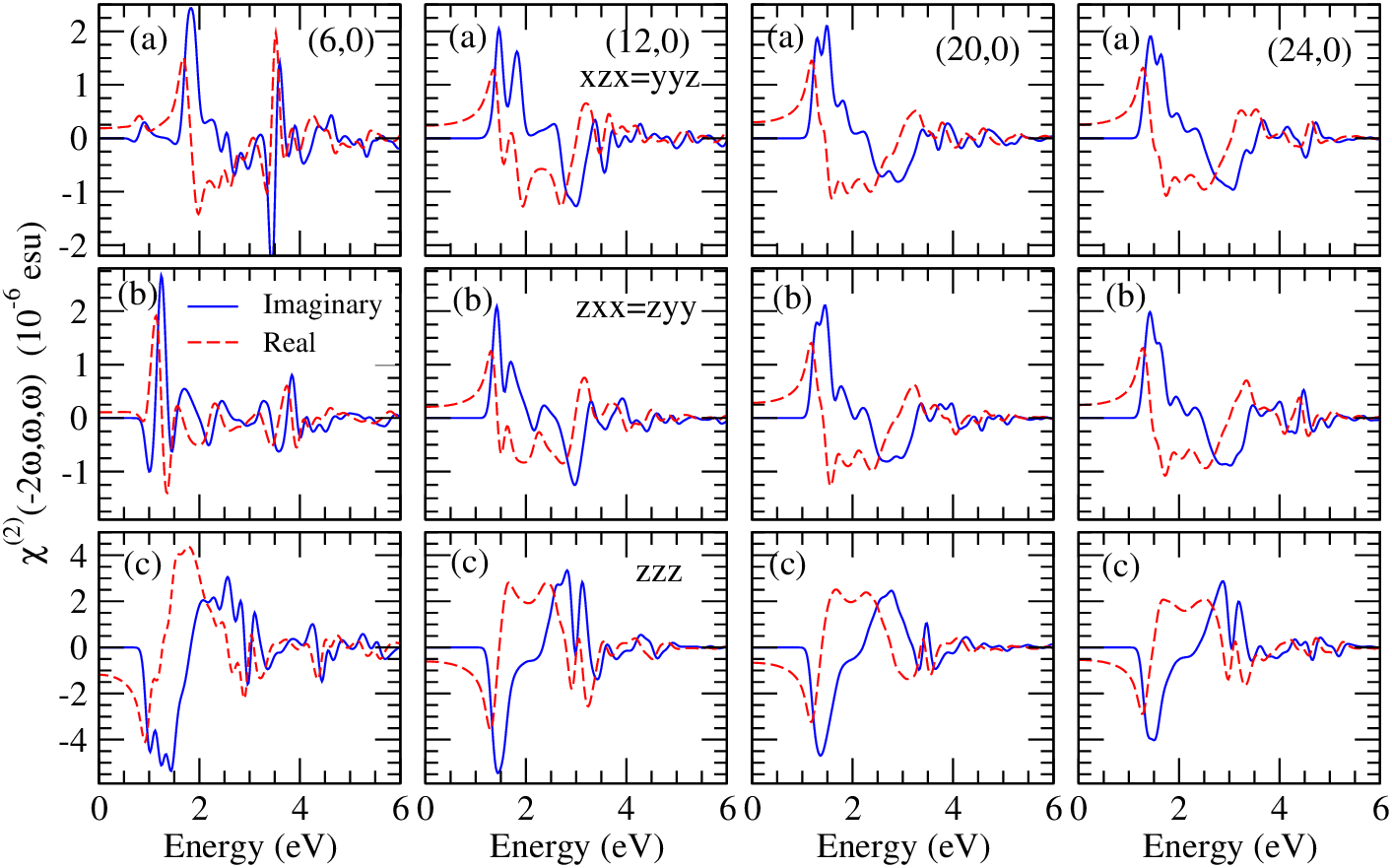}
\caption{\label{fig5} (Color online) Real (dotted line) and imaginary parts (solid line)
of $\chi^{(2)}(-2\omega,\omega,\omega)$ of the zigzag (6,0), (12,0),
(20,0) and (24,0) SiC nanotubes.}
\end{figure*}

We display in Fig. 5 the calculated real and imaginary parts of the second-order
optical susceptibility $\chi^{(2)}(-2\omega,\omega,\omega)$ for the
four selected zigzag nanotubes [(6,0), (12,0), (20,0), and (24,0)].
As for the single SiC sheet, the second harmonic generation coefficients
$\chi^{(2)}(-2\omega,\omega,\omega)$ for the zigzag SiC-NTs are significant in the entire
range of the optical photon energy ($\hbar\omega$). Indeed, they can be 
up to ten times larger than that of bulk SiC polytypes (see Figs. 1 and 5).
Moreover, for the photon energy smaller than 2.0 eV, the $\chi^{(2)}$
is purely dispersive (i.e., real and lossless) (Fig. 5). 
Interestingly, the magnitude of $\chi^{(2)}_{zzz}$ is the largest (Fig. 5).
For $\chi^{(2)}_{zzz}$, the electric field of both the incoming and outgoing
photons is polarized parallel to the tube axis and thus the electric depolarization
effect would be essentially zero (see Ref. \onlinecite{guo07} and references therein). 
This may be particularly important for nonlinear optical applications. 
The prominant features in each nonzero component of $\chi^{(2)}$ for all the
zigzag SiC-NTs except the small-diameter (6,0) SiC-NT, look rather similar (Fig. 5).
In particular, the spectra of the three $xzx$, $zxx$ and $zzz$ components of $\chi^{(2)}$ 
for the (20,0) and (24,0) SiC-NTs are nearly identical (Fig. 5). 
We also note that the shape of the spectra of $\chi^{(2)}_{zzz}$ and $\chi^{(2)}_{zxx}$
for all the zigzag SiC-NTs look very similar, except the difference in sign.
For the zigzag SiC-NTs with a larger diameter such as (12,0), (20,0) and (24,0), 
the spectrum of $\chi^{(2)}_{xzx}$ is also similar to that of $\chi^{(2)}_{zxx}$ (Fig. 5).
The magnitude of the $\chi^{(2)}_{xzx}$ and $\chi^{(2)}_{zzz}$ components decreases slightly 
as the diameter of the tubes increases [e.g., from (6,0) to (24,0)], while the magnitude
of the $\chi^{(2)}_{zxx}$ remains more or less independent of the diameter (Fig. 5). 
Furthermore, both the magnitude and shape
of the $\chi^{(2)}_{zzz}$ spectrum for the zigzag SiC-NTs with a larger
diameter, e.g., (20,0) and (24,0), approach to that of the single SiC sheet (Fig. 4), as they should. 

\begin{figure}[tb]
\includegraphics[width=8cm]{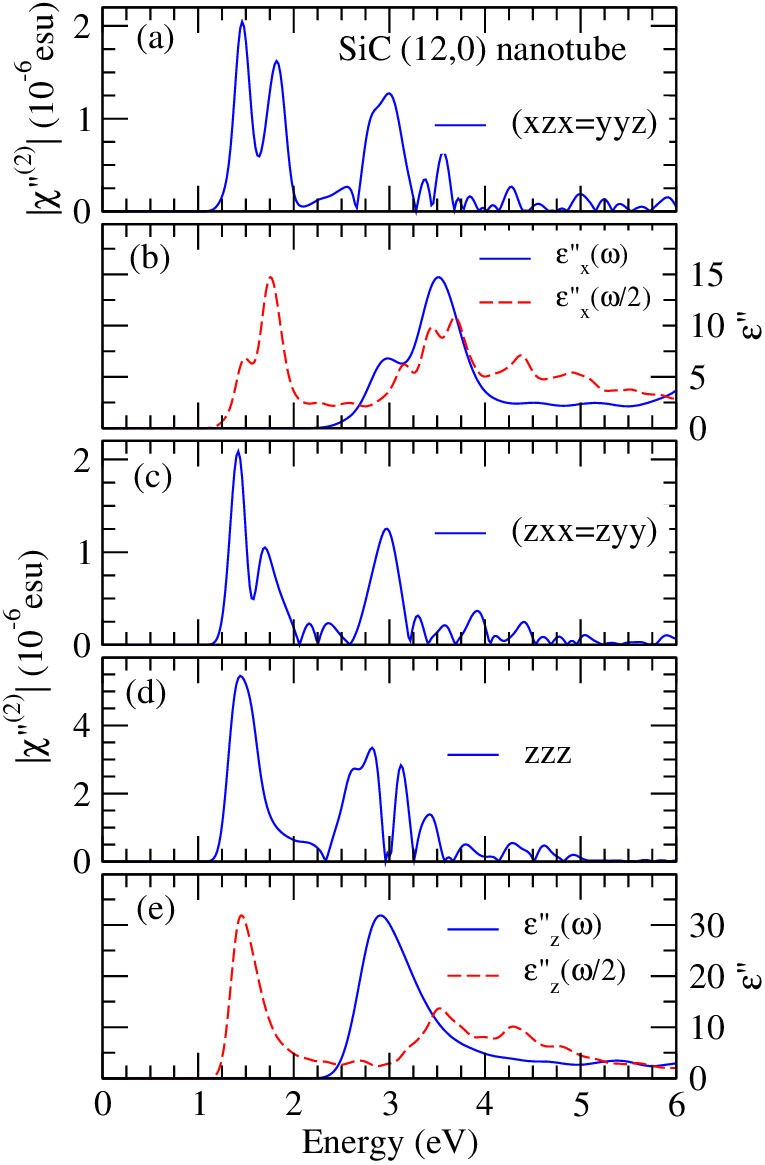}
\caption{\label{fig6} (Color online) Absolute value of the imaginary part 
of $\chi^{(2)}(-2\omega,\omega,\omega)$ (a, c, d) as well as
$\varepsilon''(\omega)$ and $\varepsilon''(\omega/2)$ (imaginary part
of the dielectric function) (b, e) from Ref. ~\cite{ijwu} of the zigzag (12,0)
SiC nanotube.}
\end{figure}

As for the single SiC sheet, in order to understand the features in the 
calculated $\chi^{(2)}$ spectra of the zigzag SiC-NTs, the absolute values of the imaginary part
$\chi''^{(2)}$ of all the nonzero components of the (12,0) SiC-NT are plotted and compared with the
absorptive part of the dielectric function
$\varepsilon''$ from our previous publication~\cite{ijwu} in Fig. 6.  
Strikingly, the first prominent peak between 1.0 and 2.0 eV 
in the $\chi_{zzz}''^{(2)}$ spectrum is almost identical to the first peak in
the $\varepsilon_z''(\omega/2)$ (see Fig. 6(d) and 6(e)), indicating that it
is due to two-photon resonances. In contrast, the second peak between 2.5 and 3.0 eV
in the $\chi_{zzz}''^{(2)}$ spectrum is very similar to the
first peak in the $\varepsilon_z''(\omega)$, suggesting that 
it is caused by the single-photon resonances. Nevertheless, 
both these single- and double-photon resonances
involve only interband $\pi \rightarrow \pi^{\ast}$
and $\sigma \rightarrow \sigma^{\ast}$ optical transitions for the
electric field vector ${\bf E}$ polarized parallel to the tube axis 
($E \parallel \hat{z}$)~\cite{ijwu}.  
Fig. 6 also shows that the double-peak structure between 1.0 and 2.0 eV in
both the $\chi_{xzx}''^{(2)}$ and $\chi_{zxx}''^{(2)}$ spectra is mainly due to the
two-photon resonances with $E\perp \hat{z}$ [cf. $\varepsilon_x''(\omega/2)$]
(see Fig. 6(a)-6(c)), while, in contrast, the second feature in the photon energies above
3.0 eV perhaps comes predominantly from the single-photon resonances
for $E\perp \hat{z}$ [cf. $\varepsilon_x''(\omega)$]. This conclusion is further 
supported by the fact that the magnitude of $\varepsilon_x''(\omega)$ ($\varepsilon_x''(\omega/2)$)
is only about half of that of $\varepsilon_z''(\omega)$ ($\varepsilon_z''(\omega/2)$),
and concurrently the magnitude of $\chi_{xzx}''^{(2)}$ and $\chi_{zxx}''^{(2)}$ 
is only about half of that of $\chi_{zzz}''^{(2)}$ too. 

\begin{figure*}[tb]
\includegraphics[width=16cm]{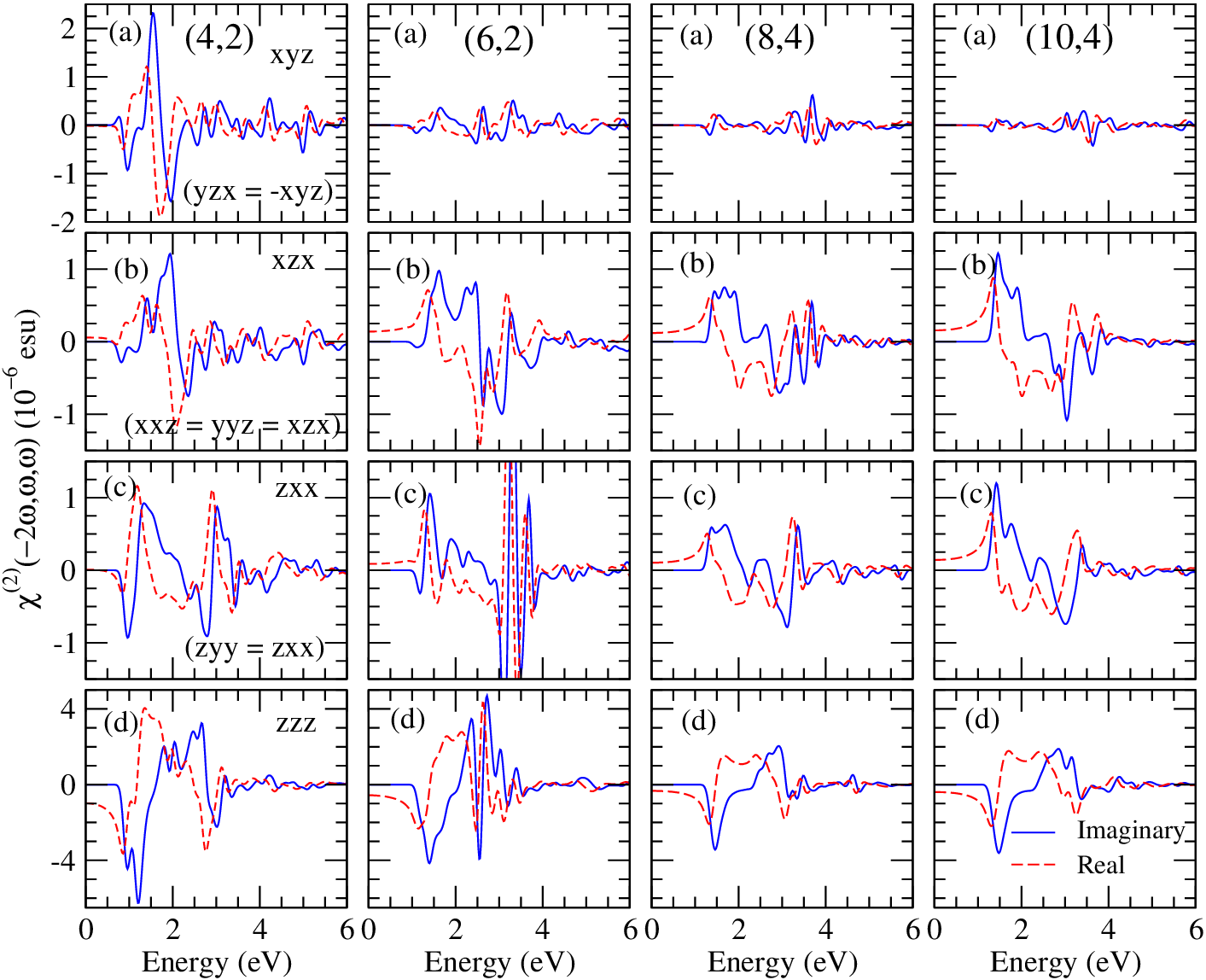}
\caption{\label{fig7} (Color online) Real (dotted line) and imaginary parts (solid line)
of $\chi^{(2)}(-2\omega,\omega,\omega)$ of the chiral (4,2), (6,2),
(8,4) and (10,4) SiC nanotubes.}
\end{figure*}

In Fig. 7, the calculated real and imaginary parts of the second-order
optical susceptibility $\chi^{(2)}(-2\omega,\omega,\omega)$ for all
the four chiral nanotubes [(4,2), (6,2), (8,4) and (10,4)] are displayed.
In general, the spectra of each component
of the second-order optical susceptibility for the chiral SiC-NTs with a moderate
diameter [e.g., (8,4) and (10,4)] are similar.
For all the chiral SiC-NTs except the ultra-small diameter (4,2) SiC-NT, 
remarkably, the $zzz$ component is nearly four times
larger than all the other nonvanishing components. Another common feature is that both the real
and imaginary parts of $\chi^{(2)}(-2\omega,\omega,\omega)$ show a
rather oscillatory behavior, particularly for the $xyz$ and $xzx$ components
(Fig. 7). The amplitude of these oscillatory real and imaginary
parts is rather large in the photon energies of 1.0$\sim$ 4.0 eV
for the chiral SiC-NTs with a small diameter. It should also be noted that 
the shape and magnitude of the $xzx$ and $zxx$ components look very much alike
(Fig. 7(b)-7(c)). Finally, both the magnitude and shape
of the $\chi^{(2)}_{zzz}$ spectrum for the chiral SiC-NTs with a larger
diameter, e.g., (10,4), approach to that of the single SiC sheet (Fig. 4), as expected.

\begin{figure}[tb]
\includegraphics[width=8cm]{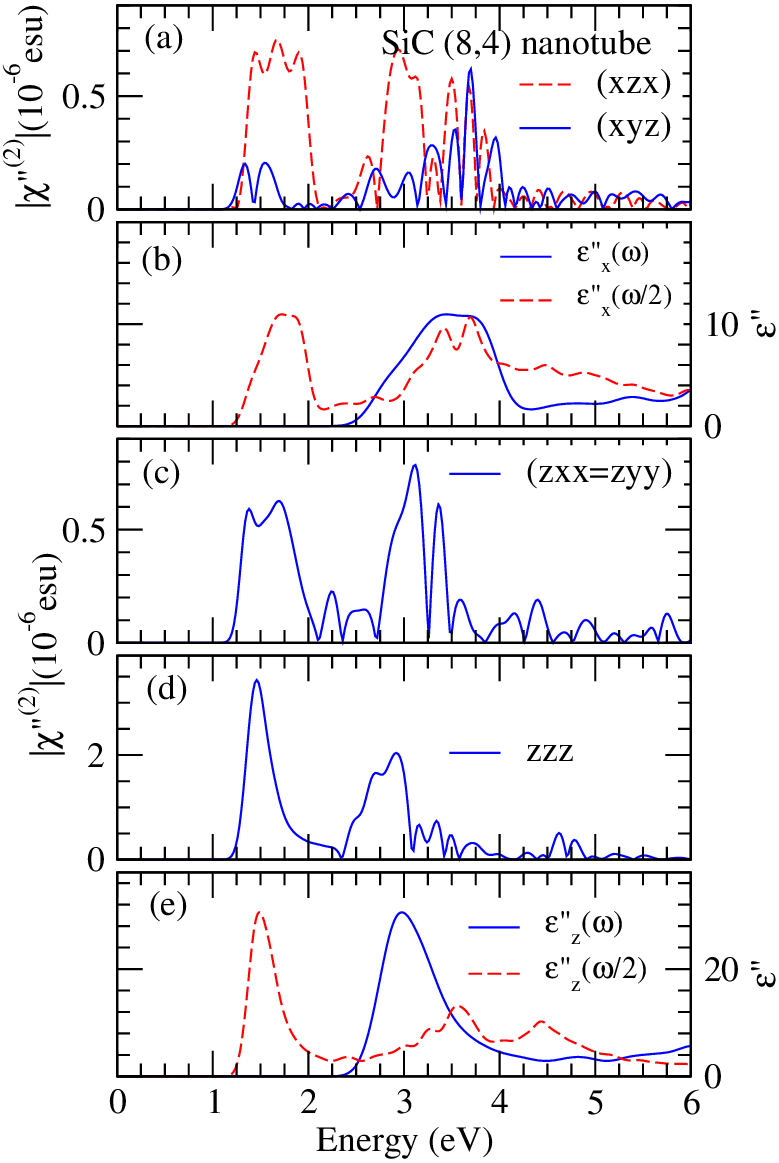}
\caption{\label{fig8} (Color online) Absolute value of the imaginary part
of $\chi^{(2)}(-2\omega,\omega,\omega)$ (a, c, d) and 
$\varepsilon''(\omega)$ and $\varepsilon''(\omega/2)$ (imaginary part
of the dielectric function) (b, e) from Ref. ~\cite{ijwu} of the zigzag (8,4)
SiC nanotube.}
\end{figure}

Again, in order to understand the structures in the
calculated $\chi^{(2)}$ spectra of the chiral SiC-NTs, the absolute values of the imaginary part
$\chi''^{(2)}$ of all the nonzero components of the (8,4) SiC-NT are plotted and compared with the
absorptive part of the corresponding dielectric function
$\varepsilon''$ from our previous publication~\cite{ijwu} in Fig. 8.
Remarkably, the first prominent peak between 1.0 and 2.0 eV
in the $\chi_{zzz}''^{(2)}$ spectrum is almost identical to the first peak in
the $\varepsilon_z''(\omega/2)$ (see Fig. 8(d)-8(e)), indicating that it
is due to two-photon resonances. On the other hand, the second structure between 2.5 and 3.5 eV
in the $\chi_{zzz}''^{(2)}$ spectrum may be correlated with the
first peak in the $\varepsilon_z''(\omega)$, suggesting that
it is caused by the single-photon resonances. As for the zigzag SiC-NTs,
both these single-photon and double-photon resonances
involve only interband $\pi \rightarrow \pi^{\ast}$
and $\sigma \rightarrow \sigma^{\ast}$ optical transitions for the
electric field vector ${\bf E}$ polarized parallel to the tube axis
($E \parallel \hat{z}$)~\cite{ijwu}.
Fig. 8 further suggests that the feature between 1.0 and 2.0 eV in
the spectra of both the $\chi_{xzx}''^{(2)}$ and $\chi_{zxx}''^{(2)}$ as well 
as $\chi_{xyz}''^{(2)}$ may be attributed to the
two-photon resonances with $E\perp \hat{z}$ [cf. $\varepsilon_x''(\omega/2)$]
(see Fig. 8(a)-8(c)), while, in contrast, the second feature in the photon energies above
3.0 eV is mainly due to the single-photon resonances
for $E\perp \hat{z}$ [cf. $\varepsilon_x''(\omega)$]. 

\subsection{Linear electro-optical coefficient}

\begin{table*}
\caption{Calculated static refraction index $n$, second-order optical
susceptibility $\chi^{(2)}$ and linear electro-optical 
coefficient $r_{abc}$ of the zigzag and chiral SiC nanotubes.}
\begin{ruledtabular}

\begin{tabular}{c c c c}
      & $n_{x}$ ($n_{z}$) 
      & $\chi^{(2)}_{xzx}$, $\chi^{(2)}_{zxx}$, $\chi^{(2)}_{zzz}$ 
      & $r_{xzx}$, $r_{zxx}$, $r_{zzz}$ \\ 
      &     &  (pm/V) & (pm/V) \\ \hline
 (5,0)  & 2.50 (3.39) &  0.0,  0.0, 0.0       & 0.0, 0.0, 0.0 \\
 (6,0)  & 2.53 (3.31) & 77.3, 45.5, -495.5 & -2.19, -1.29, 8.21 \\
 (8,0)  & 2.56 (3.23) & 87.6, 68.0, -323.1 & -2.56, -1.98, 5.92 \\
 (9,0)  & 2.58 (3.23) &  0.0,  0.0, 0.0       &  0.0,  0.0, 0.0 \\
 (12,0) & 2.60 (3.21) & 98.9, 88.1, -255.4 & -2.84, -2.53, 4.82 \\
 (16,0) & 2.62 (3.21) & 104.7, 96.9, -241.9 & -2.96, -2.74, 4.57 \\
 (20,0) & 2.68 (3.27) & 125.0, 119.6, -279.2 & -3.25, -3.11, 4.85 \\
 (24,0) & 2.63 (3.19) & 107.1, 102.1, -227.7 & -3.05, -2.91, 4.39 \\
 (4,2)  & 2.74 (3.37)  & 23.9, 4.46,  -410.4 & -0.56, -0.11, 6.36   \\
 (6,2)  & 2.56 (3.16)  & 58.2, 37.0, -236.6  & -1.78, -1.13, 4.75   \\
 (8,4)  & 2.62 (3.17)  & 50.3, 43.6, -138.6 & -1.46, -1.26, 2.75  \\
 (10,4) & 2.62 (3.17)  & 65.3, 59.3, -164.3  & -1.89, -1.72, 3.25   \\
\end{tabular}
\end{ruledtabular}
\end{table*}

We list in Table IV the calculated zero frequency linear electro-optic coefficient
$r(0)$ as well as the corresponding second-order nonlinear optical susceptibility
$\chi^{(2)}(0,0,0)$ of the SiC-NTs. The $r(0)$ is calculated from the 
corresponding $\chi^{(2)}(0,0,0)$ by using Eq. (7). 
The refraction index $n(0) (=\sqrt{\varepsilon(0)})$
is derived from the calculated static dielectric constant
$\varepsilon(0)$ which has been reported in our recent publication~\cite{ijwu}.
Table IV shows that apart from the (5,0) and (9,0) SiC-NTs which
have no nonvanishing $r_{abc}(0)$, and also the small diameter 
(6,0) and (8,0) SiC-NTs, all the other zigzag SiC-NTs have very similar
linear electro-optical coefficients, as for the static dielectric constant
and polarizability~\cite{ijwu}. Nevertheless, $r_{zzz}(0)$ decreases
slightly as the diameter increases, while $r_{xzx}(0)$ and $r_{zxx}(0)$ increase
slightly as the diameter goes up. Remarkably, the static $\chi_{zzz}^{(2)}$
for the ultra-small diameter (6,0) is very large (Table IV), being nearly thirteen times
larger than the largest component of $\chi^{(2)}$ of all the four
bulk SiC polytypes (Table II). This suggests that the SiC-NTs would be useful 
for applications in the non-linear optical devices. Nonetheless, the $\chi_{zzz}^{(2)}$
decreases rapidly as the diamter of the zigzag SiC-NTs increases, and,
for the large diamter (24,0) SiC-NT, approaches the value of $\chi_{aab}^{(2)}$ of 
the single SiC sheet (Table II). The magnitude of the $\chi_{aab}^{(2)}$ 
is about half of the magnitude of the $\chi_{zzz}^{(2)}$ for the (6,0) SiC-NT. 
We note that $\chi_{xzx}^{(2)}$ and $\chi_{zxx}^{(2)}$
are rather similar, and their magnitudes increase as their tube diameters
increases. Finally, for the larger diameter zigzag SiC-NTs,
the $\chi_{zzz}^{(2)}$ is about two times larger than $\chi_{xzx}^{(2)}$ 
and $\chi_{zxx}^{(2)}$ (Table IV).
 
The chiral SiC-NTs have two additional nonvanishing components $\chi_{xyz}^{(2)}$
and $\chi_{yzx}^{(2)}$. Nevertheless, the calculated static
values of $\chi^{(2)}_{xyz}$ and $\chi^{(2)}_{yzx}$ are negligibly small
(i.e., within the numerical uncertainty), thereby satisfying
the requirement by the so-called Kleinman symmetry~\cite{kle62} which
demands that $\chi^{(2)}_{xyz}(0)  = \chi^{(2)}_{yzx}(0)$. 
Consequently, the corresponding static
linear electro-optical coefficients $r_{xyz}(0)$ and $r_{yzx}(0)$ are zero too.
Therefore, $\chi_{yzx}^{(2)}$, $\chi^{(2)}_{xyz}$, $r_{xyz}(0)$ and $r_{yzx}(0)$ 
for the chiral SiC-NTs are not listed in Table IV.
As for the nonvanishing static components, $r_{xzx}(0)$ and $r_{zxx}(0)$ for the
(6,2), (8,4) and (10,4) SiC-NTs are quite close. 
$\chi^{(2)}_{zzz}(0)$ and $r_{zzz}(0)$ decrease as the tube diameter increases, 
while, in contrast, 
$\chi^{(2)}_{zxx}(0)$ and $r_{zxx}(0)$ increase with the tube diameter. 
$r_{xzx}(0)$ and $r_{zxx}(0)$ for the (4,2) is somewhat smaller than that of the 
other chiral SiC-NTs. 

The zero frequency second-order nonlinear optical susceptibility
$\chi^{(2)}(0,0,0)$ of the SiC-NTs (Table IV) are generally a few times larger
than the counterparts of the corresponding BN-NTs~\cite{Lin}. 
In particular, the $\chi_{zzz}^{(2)}(0,0,0)$ of the (6,0) SiC-NT
is about thirteen times larger than that of the (6,0) BN-NT~\cite{Lin}.
On the other hand, the $\chi_{zzz}^{(2)}(0,0,0)$ of the (24,0) SiC-NT
is only about six times larger than that of the (24,0) BN-NT~\cite{Lin}. 
This suggests that compared with BN-NTs, SiC-NTs would be a better nonlinear 
optical material. However, the low frequency linear electro-optic coefficient
$r(0)$ of the SiC-NTs (Table IV) are only slightly larger than 
the counterparts of the corresponding BN-NTs~\cite{Lin}. 
For example, the $r_{zzz}(0)$ of the (6,0) and (24,0) SiC-NTs is, respectively,
1.7 and 1.3 times larger than that of the (6,0) and (24,0) BN-NTs.
This is because the refraction index $n$ of the SiC-NTs (Table IV) 
are also larger than that of the BN-NTs~\cite{Lin} [see Eq. (7)]. 


\section{Summary}

We have carried out a systematic {\it ab initio} study of
the second-order nonlinear optical properties of SiC-NTs
within density functional theory in the local density approximation.
We used the highly accurate full-potential PAW method.
The underlying atomic structure of the SiC nanotubes was
determined theoretically.
Specifically, the properties of the single-walled zigzag 
[(5,0), (6,0), (8,0), (9,0), (12,0), (16,0), 
(20,0), (24,0)], armchair [(4,4), (5,5), (8,8), (12,12), (15,15)],
and chiral [(4,2), (6,2), (8,4), (10,4)] nanotubes have been calculated.
For comparison, the second-order nonlinear optical properties 
of bulk SiC polytypes (2H-, 4H-, 6H- and 3C-SiC) and the single 
graphitic SiC sheet have also been calculated.
Interestingly, we find that the $\chi^{(2)}_{aab}$, $\chi^{(2)}_{baa}$ and
$\chi^{(2)}_{bbb}$ for the isolated SiC sheet are large and
generally several times larger than that of the SiC polytypes.
Unlike carbon nanotubes, both the chiral and zigzag SiC-NTs have
pronounced second-harmonic generation and linear electro-optical coefficients
which are comparable to or even larger than that of the single SiC sheet. 
The prominant structures in the spectra of $\chi^{(2)}(-2\omega,\omega,\omega)$
of the SiC-NTs have been related to the features in the corresponding
linear optical dielectric function $\varepsilon (\omega)$ in terms of
single-photon and double-photon resonances. 
We also find that the $\chi_{abc}^{(2)}$ and $r_{abc}$ coefficients of the SiC-NTs are 
up to thirteen-times larger than the counterparts of the corresponding BN-NTs. 
Therefore, the SiC-NTs would be promising nonlinear optical materials
for applications in, e.g., second-harmonic generation, sum frequency generation and
electro-optical switches. We hope that this work would stimulate
experimental investigations into the second-order nonlinear optical properties of the
SiC-NTs.

\section{Acknowledgments}
The authors gratefully acknowledge financial support from National Science Council
and NCTS of ROC. The author also thank the National Center for High-Performance 
Computing of ROC, and the Computer and Information Networking
Center (CINC) of National Taiwan University for providing CPU time. 
\\

\end{document}